\documentclass[prd,aps,twocolumn,floats]{revtex4}
\usepackage{graphicx}
\begin{document}

\title{Convergence and stability in numerical relativity}
\author{Gioel Calabrese, Jorge Pullin, Olivier Sarbach, Manuel Tiglio}
\affiliation{Department of Physics and Astronomy, Louisiana State
University, 202 Nicholson Hall, Baton Rouge, LA 70803-4001}

\begin{abstract}
It is often the case in numerical relativity that schemes that are
known to be convergent for well posed systems are used in evolutions
of weakly hyperbolic (WH) formulations of
Einstein's equations. Here we explicitly show that with several of the
discretizations that have been used through out the years, this
procedure leads to non-convergent schemes. 
That is, arbitrarily small
initial errors are amplified without bound when resolution is
increased, independently of the amount of numerical dissipation
introduced. The lack of convergence introduced by this instability can
be particularly subtle, in the sense that it can be
missed by several convergence tests, especially in 3+1 dimensional
codes.  We propose tests and methods to analyze convergence that may
help detect these situations.

\end{abstract}

\maketitle

Convergence is a central element of any numerical simulation.  It
refers to the property that if one refines the simulation by adding
more points to the grid, numerical errors should diminish.  In the
limit of zero spacing, they should go to zero and one should get the
exact solution.  Having a convergent code is a key element for
numerical simulations to have predictive power: although one will
always be limited in practice to a finite number of points in the
grid, one can extrapolate the results for more and more refined
simulations and have a very good approximation to the true (continuum)
results.  Codes that do not converge produce answers that, even if
they remain finite (at least for a while), have little predictive
power: there generically is no way to know if the solutions found
approximate the desired continuum solution.

In this paper we want to emphasize that discretization schemes that
yield convergent code for strongly hyperbolic (SH) systems of equations
do not necessarily do so for weakly hyperbolic (WH) or completely ill 
posed (CIP) systems. The relevance
of this observation is that several formulations of the Einstein 
evolution equations commonly used in numerical relativity, including,
for example, the 
Arnowitt--Deser--Misner (ADM) \cite{ADM} and
Baumgarte--Shapiro--Shibata--Nakamura (BSSN) \cite{BSSN} formulations
with fixed lapse and shift,  
are not SH.

We concretely prove that a simple system of linear WH equations with
constant coefficients, when discretized with either the iterated
Crank--Nicholson (ICN) with fixed number of iterations or second order
Runge--Kutta (2RK) methods, leads to unconditionally unstable codes,
even if numerical dissipation is explicitly added. This simple system
of equations is related to the equations one encounters in the WH
formulations of general relativity. It therefore strongly suggests
that these formulations should produce code that is not
convergent. The lack of convergence is of a particularly pernicious
nature, since in the WH case it might grow slower than the ``usual''
von-Neumann numerical instability. It is such that if one tests the
code with stationary solutions (for instance simulating a single black
hole) and performs convergence tests, one could easily be confused
into believing one has convergent code.

Consider a linear system of $m$ partial differential equations in $m$
variables of the form
\begin{equation}
\partial_t v = A\partial _xv \; ,
\label{pde}
\end{equation}
where $A$ is a constant $m\times m$ matrix and $v=(v_1,\ldots, v_m)^T$.
The system is SH if $A$ has real
eigenvalues and is diagonalizable, WH if it has real 
eigenvalues but is not diagonalizable, and CIP if it 
has complex eigenvalues. 

The iterated Crank--Nicholson (ICN) method is an explicit
approximation to the (implicit) Crank--Nicholson method first proposed
by Choptuik. In this proposal the number of iterations is not fixed
but, instead, might depend on the spatial gridpoint and time step
\footnote{We thank Matt Choptuik for discussions on this point.}. A
different notion of ICN is to a priori fix the number of iterations
\cite{ICN}.  Teukolsky \cite{ICN} showed that for the advective
equation, $\partial_t{v}= \partial_xv$ (which is SH), such scheme was
stable if one used $2,3,6,7,10,11,\ldots$ iterations (one iteration
corresponds to the 2RK method). This paper seeks to extend this proof
to more general systems of equations, including explicit artificial
viscosity.

A difference scheme is said to be stable with respect to a given norm
if there exist positive constants $\Delta x_0$ and $\Delta t_0$ and
$f(t)$ such that $ \| u^{n} \| \, \le \, f(t) \, \| u^{0} \| \,,
\label{stab_def} $ for $0<\Delta x\le\Delta x_0$, $0<\Delta t\le
\Delta t_0$, and $t=n\Delta t$ where $u^n$ is  the numerical 
solution at time $t$ and $u^0$ the initial data. That is, the
growth of the solution is bounded by some function of time $f(t)$ that is
{\em independent of the initial data and resolution.}  Some schemes
are only conditionally stable.  This means that a relationship between
$\Delta t$ and $\Delta x$  is needed for stability.
Through Lax's theorem, numerical stability in this 
sense is equivalent to convergence, provided the scheme is consistent.

Working in Fourier space and writing explicitly the solution at time
$t$ in terms of the amplification matrix $Q$, $ \hat{u}^n = Q^n
\hat{u}^0 $ yields a necessary condition for stability: the
von-Neumann condition (VNC).  It states that, for the cases considered
in this paper, the spectral radius $\rho (Q)$ of $Q$ (the maximum
eigenvalue in norm) is not greater than one. If this condition is not
satisfied, a numerical instability that grows like $\exp{(n \mu)}$,
with $\mu>0$ a constant, is present. A well known example of this
instability is the forward in time, centered in space scheme for the
advective equation. It is sometimes thought that the VNC is not only
necessary but also sufficient for stability. This often leads to
(wrongly) concluding that a discretization for a WH system is stable
based on an discrete analysis for the advection or wave equation. The
VNC is sufficient only if $Q$ can be diagonalized by a similarity
transformation whose norm, and that of its inverse, has an upper bound
that is independent of resolution \cite{kreiss}. An observation that
will be important for the stability analysis below is that, for the
cases we are discussing, $Q$ is diagonalizable if and only if $A$
is. Therefore, in the WH case the VNC will not be sufficient for
stability.

On the other hand, if the system is SH $A$ is diagonalizable (with real
eigenvalues) and so is $Q$. Working in the diagonal basis we then 
have an uncoupled system of $m$
equations for the $m$ grid functions. Repeating Teukolsky's analysis
for the ICN scheme, this system is stable if $\rho (A) \lambda \le 2$
($\lambda = \Delta t /\Delta x$ is the Courant
factor) and the number of iterations is $p = 2,3,6,7,10,11,\ldots$, if no
explicit dissipation is added. In evolutions of
SH equations, the addition of dissipation may help stabilize schemes
that are otherwise unstable. For example, the VNC for the 2RK case
with third order dissipation is 
\begin{equation}
(\rho (A)\lambda)^4 \leq 8\lambda \tilde{\sigma}\leq 1 \; , \label{vnc}
\end{equation}
where
$\tilde{\sigma }$ is the dissipation parameter.  Since $Q$
is diagonal, this condition is necessary and sufficient for
stability. 

Below we will explicitly prove that for CIP or WH cases the ICN and
2RK methods are unstable, independently of the amount of dissipation,
number of iterations and Courant factor.  Although our results are
only proven for a system of linear equations with constant
coefficients, they are significant in several ways for numerical
relativity. First of all, if one considers situations that are small
perturbations off Minkowski spacetime, then a system of equations
identical to the one we considered {\em does} appear in the linearized
Einstein equations written in first order form in space and time if
the system of evolution equations used is WH or CIP. Since far away
from binary black holes the spacetime is approximately flat, this
situation does arise in realistic simulations. Could non-linearities
change our conclusions?  It is possible, but unlikely. Usually when
instabilities are established for linear systems the addition of
non-linear terms only makes instabilities worse \cite{kreiss}.

The type of lack of convergence differs in the CIP and WH case. In the
CIP case the VNC is easily violated, codes crash very fast and
non-convergence is easy to spot. In the WH case the VNC may be
satisfied and crashes may occur very late (if they occur at all).  If
one performs routine convergence analyses in the region before the
crash one might appear to see convergence, especially if the initial
data has few (and low) frequency components, unless one increases
enough the resolution or runs for long enough time.

We now show that these different types of lack of convergence do
appear in numerical relativity simulations, even in very simple 1D
ones with periodic boundary conditions. We have performed non-linear
evolutions of a plane gauge wave spacetime (that is, the spacetime is
flat, though in a time-dependent slicing),
\begin{equation}
ds^2 = e^{A\sin{(t+x)}}(-dt^2 + dx^2) + dy^2 + dz^2\;.
\end{equation}

\begin{figure}[ht] 
\begin{center} 
\includegraphics*[height=5.5cm]{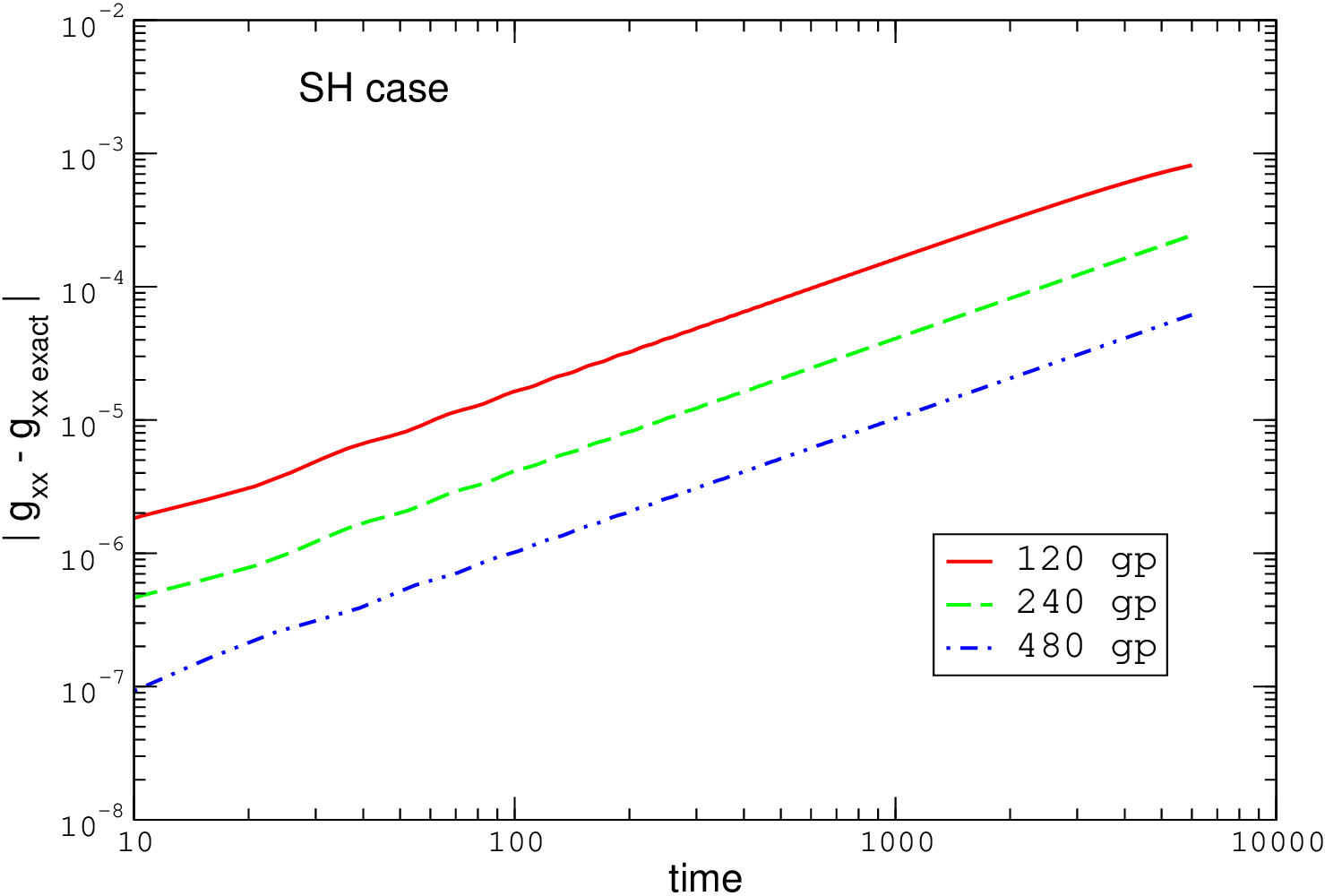}
\includegraphics*[height=5.5cm]{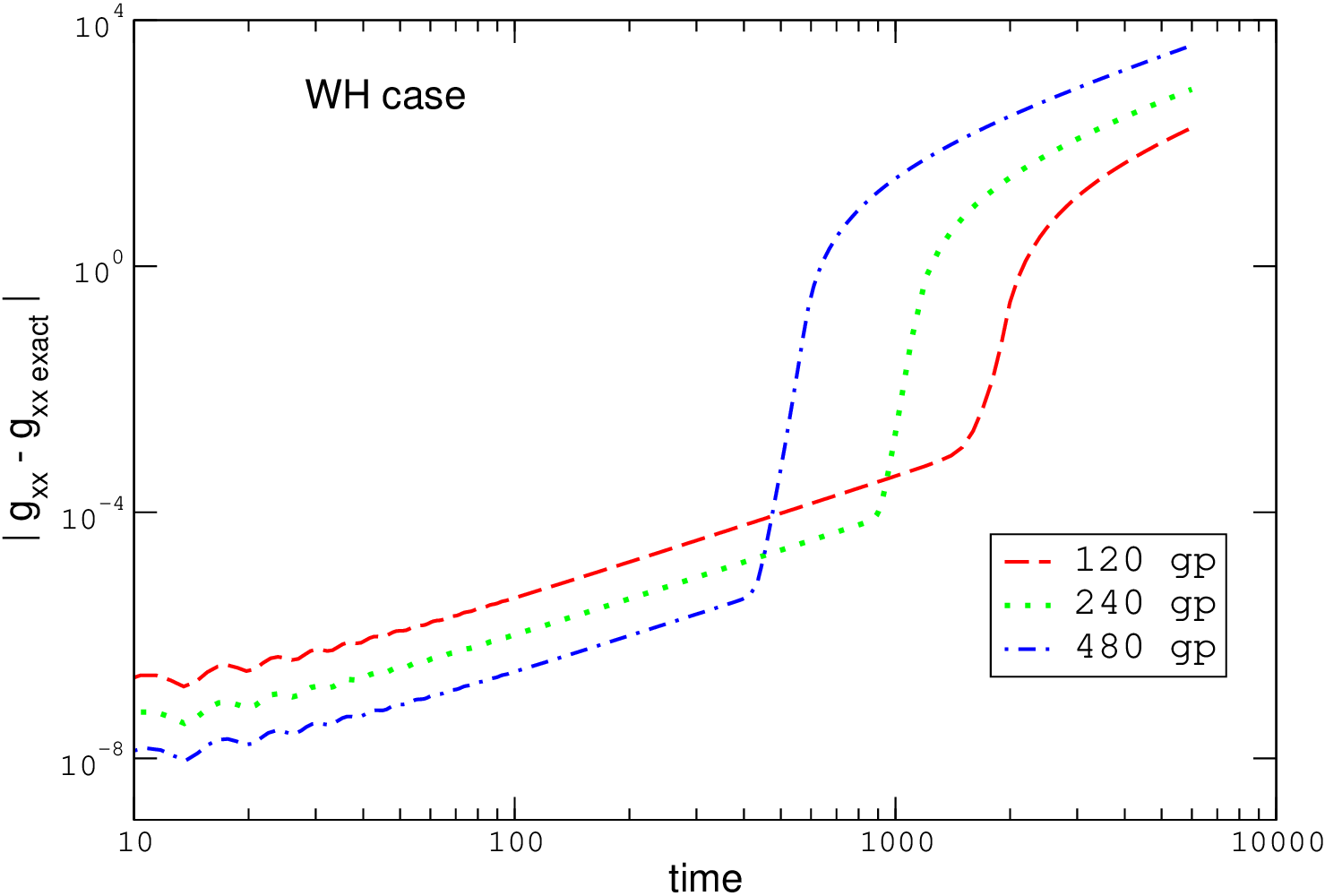}
\includegraphics*[height=5.5cm]{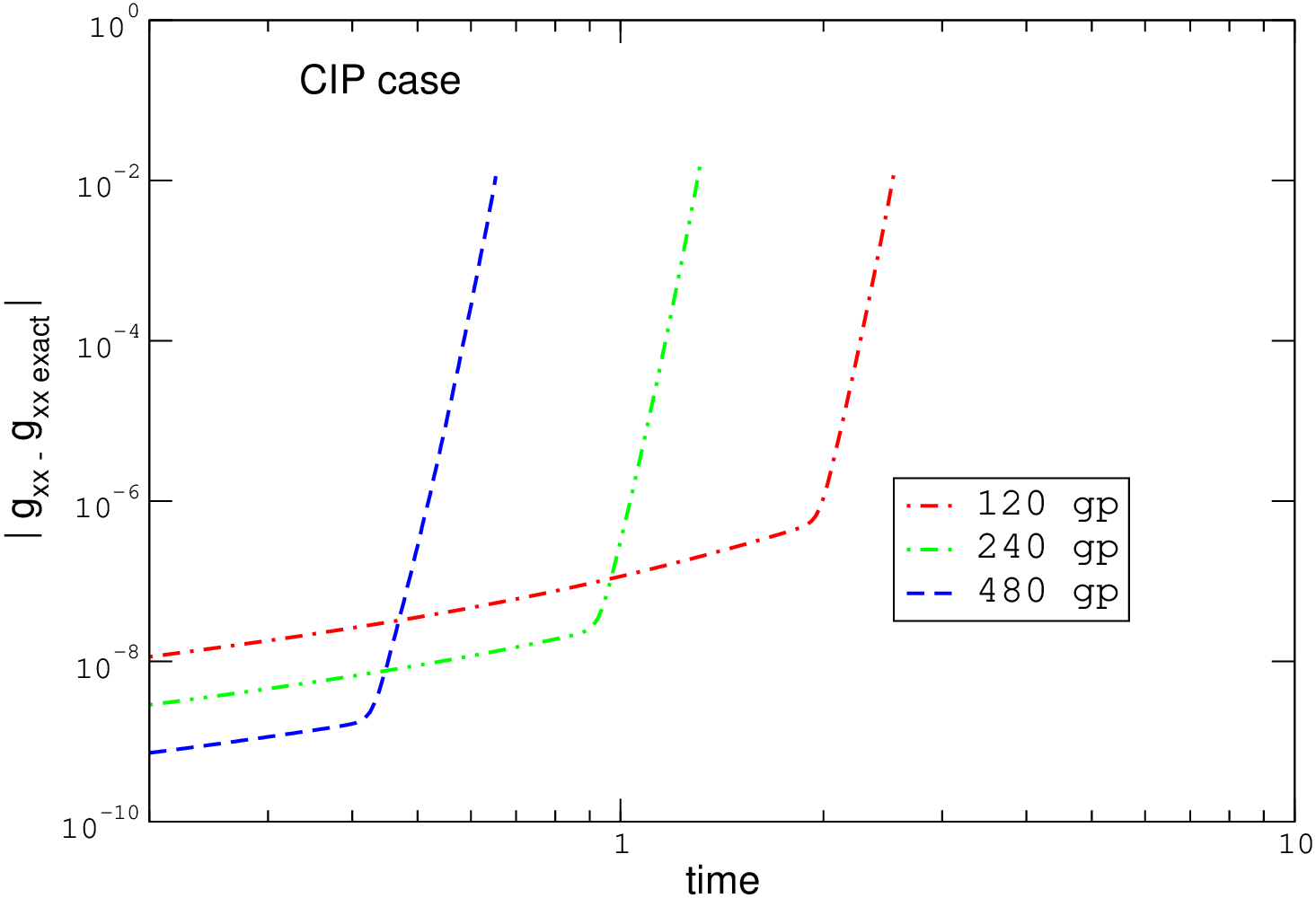}
\caption{$L_2$ norm of the errors for the metric in the SH, WH and 
CIP cases. \label{errors}}
\end{center}
\end{figure}

Figure \ref{errors} shows results of an evolution using 2RK with
$\sigma =1/2, 0,-1/2$ in the Einstein--Christoffel (EC) system
\cite{ec} of evolution equations. The latter is a symmetric-hyperbolic
reformulation of Einstein's evolution equations that includes a
parameter $\sigma$ that densitizes the lapse. In the EC formulation,
$\sigma =1/2$, but by tuning this parameter we can make the system WH
($\sigma =0$) or CIP ($\sigma =-1/2$). The whole $30$ equations of the
EC system are evolved, but all quantities are assumed to depend only
on $t$ and $x\in[-\pi, \pi]$. The Courant factor is set to $1/2$, the
dissipation parameter to $\tilde{\sigma}=0.02$, and the evolution is
followed for $1,000$ crossing times. In the SH case the code is
convergent for all resolutions tested, see \cite{stability} for a
detailed discussion.  On the other hand, in the same code with $\sigma
=-1/2$ a lack of convergence becomes apparent immediately (before one
crossing time), the errors become bigger and the code crashes earlier
as resolution is increased. In the WH case the code is also not
convergent but the effect is less noticeable, since with the chosen
values of dissipation and Courant factor the VNC, Eq.(\ref{vnc}), is
satisfied.
\begin{figure}[ht] 
\begin{center} 
\includegraphics*[height=5.5cm]{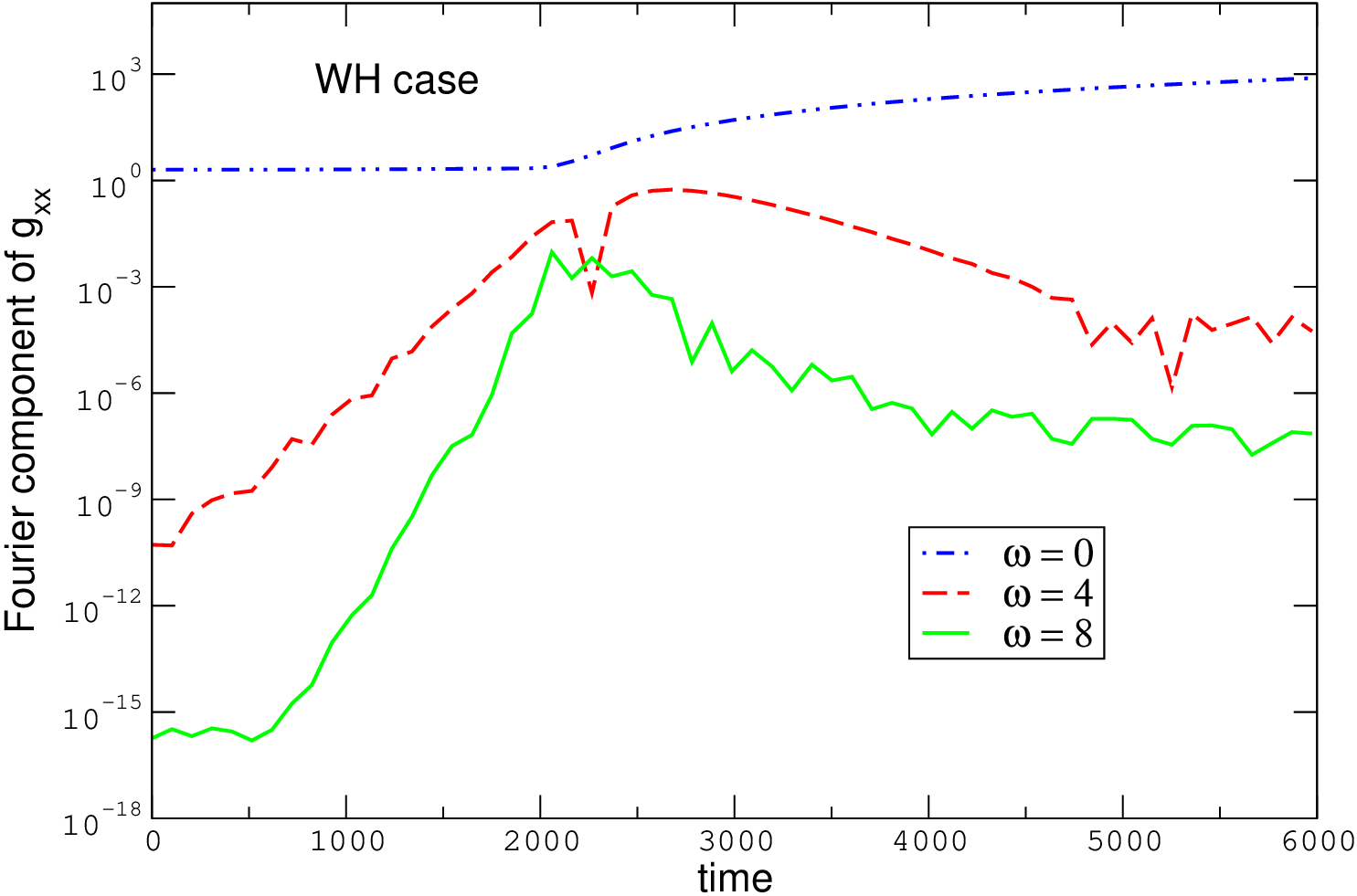}
\caption{The lack of convergence in the particular WH case considered
is triggered by non-zero
frequency modes that are suppressed in the initial data, but grow
exponentially. Notice that after the lack of convergence becomes
 obvious (at $t=2000$ for
this resolution) no further exponential growth takes place and the
code does not crash, in the example we are considering. }
\label{Fourier}
\end{center}
\end{figure} 

If one performed only a few runs, with, say $120$ and $240$ gridpoints
as is typical in 3D convergence studies, one would have to wait for
around $150$ crossing times in order for the lack of convergence in
this WH example to become obvious. To put these numbers in context,
suppose one had a similar situation in a 3D black hole
evolution. Suppose the singularity is excised, with the inner boundary
at $r=M$ and the outer boundary at $20M$.  In this case $120$ and
$240$ gridpoints correspond to grid spacings of, approximately, $M/5$
and $M/10$, respectively. If one had to evolve up to $150$ crossing
times in order to notice the lack of convergence, that would
correspond to $t\approx 3000M$, which is more than what present 3D
evolutions typically last. One could therefore be misled to think
that the code is convergent. Repeating the same runs with an
initial data with more frequencies (say, a non-stationary black hole)
would make the instability manifest in a shorter timescale. By making
a spatial Fourier decomposition (Figure \ref{Fourier}) of the
numerical solution, we have found that in the WH and CIP cases there
are always non-zero frequency modes growing exponentially from the
very beginning, though sometimes starting at truncation error. By
performing such decomposition one can detect that the code is not
convergent much before this becomes obvious in the overall errors.

We have done simulations with different spectral distributions on the
initial data, different number of iterations in the ICN method,
different Courant factors, and different values of dissipation.  The
time and resolution at which the numerical instability becomes obvious
depends on all these factors, but lack of convergence is always
present in the CIP and WH cases, while the SH runs do converge.  Too
much dissipation violates the VNC, resulting in a more severe
instability and this is immediately seen in numerical experiments, see
\cite{stability} for details.

We have also found similar results performing simulations with the
same initial data but with Kidder-Scheel-Teukolsky's many
parameter family of formulations of Einstein's equations \cite{kst},
leaving all parameters fixed and changing only one of them at a time,
achieving different levels of hyperbolicity. In particular, we have
found lack of convergence in the ADM equations as well (which is a
subsytem of this family).

The results presented in this paper do not represent a special feature
of the ICN method but instead, only reflect the fact that the
definition of numerical stability is just a discrete version of well
posedness. Therefore, difference schemes approximating ill posed
problems can be expected to be non-convergent.  In the absence of
boundaries this is the case for CIP or, generically, WH
formlations. If there are boundaries, strong hyperbolicity is not
enough and extra care has to be taken in order to guarantee well
posedness; wrong boundary conditions not only lead to inconsistencies
but also to lack of convergence (see, e.g., \cite{tadmor}).

Although it is not possible to prove in a definitive way that a code
is convergent only by numerical experiments, the previous examples
suggest some of the pathologies that one should look for. Namely,
Fourier modes that are not convergent but that are very small for a
while, since they are initially suppressed. Also, notice from the WH
simulations above shown that a code does not need to crash nor exhibit
violent growth in order not to converge. The main lesson learned from
this paper is that one should exercise significant care in numerical
simulations before empirically concluding that the simulation is
convergent, especially if the formulation of Einstein's equations used
is not SH or its level of hyperbolicity is unknown.

We wish to thank D. Arnold, M. Choptuik, P. Laguna, L. Lehner,
M. Miller, R. Price, O. Reula, and S. Teukolsky for comments.  This
work was supported in part by NSF grant PHY9800973, the Horace Hearne
Jr. Institute for Theoretical Physics, the Swiss National Science
Foundation, and Fundaci\'on Antorchas.

{\em Proof of non-convergence for CIP and WH cases:} A usual way of 
discretizing the right hand side of Eq.(\ref{pde}) is
using centered differences plus third order explicit dissipation. By
this one means solving $\partial_t v=Cv/\Delta t$, with $ C = A\lambda
\delta_0 /2 - \tilde{\sigma}\lambda \delta^4 $, where $I$ is the
identity matrix, $\delta_0 v_k =v_{k+1}-v_{k-1}$, and $\delta^4 v_k =
v_{k+2}-4v_{k+1} + 6v_k -4v_{k-1} +v_{k-2}$. 
 This is equivalent to discretizing $\partial_t v = A\partial
_xv - I\tilde{\sigma }(\Delta x)^3\partial_x^4v$ using centered
differences (using first order dissipation in our proof,
i.e.~discretizing $\partial_t v = A\partial
_xv + \tilde{\sigma }\Delta x\partial_x^2v$, gives similar results).

The ICN method with $p$ iterations for the partial differential
 equation (\ref{pde}) can be written as 
\begin{equation}
v^{n+1}_k = \left(1+
2 \sum_{j=1}^{p+1} \frac{C^j}{2^{j}}
\right)v^n_k \label{cn} \; ,
\end{equation}
where $v_k^n= (v_{1\,k}^n, \,\ldots, v_{m\,k}^n)^T$. The index
$n$ corresponds to the time step and $k$ to the spatial mesh point.
Since we are considering an initial value problem on all of $
{\sf R\hspace*{-0.9ex}\rule{0.15ex}{1.5ex}\hspace*{0.9ex}}
$,
$-\infty<k<+\infty$ (our proof can be  easily modified for periodic
boundary conditions, and the results are the same). 

In order to analyze stability we work in the basis in which $A$
takes its canonical form. That is, one multiplies both sides of
Eq.(\ref{cn}) by $T$, where $TAT^{-1}=J$ has the canonical Jordan form, and
analyzes the equation 
\begin{equation}
u^{n+1}_k = \left(1+
2 \sum_{j=1}^{p+1} \frac{L^j}{2^{j}}
\right)u^n_k \label{CN} \; ,
\end{equation}
where $L = TCT^{-1}= J\lambda \delta_0 /2 - I\tilde{\sigma}\lambda \delta^4$. 
Any conclusions regarding stability will hold as well for the original
variable $v$. Working in Fourier space, $u_k = \int_{-\pi }^{\pi
}\exp{(ik\xi)}\hat{u}(\xi) d\xi$, 
$
Q =  1+2 \sum_{j=1}^{p+1} {\hat{L}^j}/{2^{j}}
$
and $\hat{L} = J\lambda i\sin(\xi) - 16\tilde{\sigma}\lambda
\sin^4(\xi/2) $. 

If the system is CIP, $J$ has at least one complex eigenvalue $c$,
say $J_{11} = c$. In this case,
$\hat{L}_{11} = c\lambda i\sin(\xi) - 16\tilde{\sigma}\lambda
\sin^4(\xi/2)$, and the eigenvalue $Q_{11}$ has norm $1  - 
\lambda Im(c)\xi +O(\xi ^2)$.
Therefore, the VNC is violated for sufficiently small $\xi $ with the
appropriate sign and 
the scheme is unconditionally unstable, as expected.

If the system is WH then $J$
has at least one Jordan block and so does $Q$. In this basis
all the Jordan blocks are
uncoupled and we can consider one at a time. We assume there
is one block of dimension $2\times 2$ (the proof actually holds for
higher dimensionality as well). The canonical form for such a block,
the resulting amplification matrix, and its n-th power are
$$
J=
\left(
\begin{array}{cc}
c & 1 \\
0 & c
\end{array}
\right) ,\, Q=
\left(
\begin{array}{cc}
a & b \\
0 & a
\end{array}
\right)
,\,
Q^n = \left(\begin{array}{cc}
a^n & a^{n-1}nb \\
0 & a^n 
\end{array}\right)
\;.
$$
Suppose the VNC $|a|\leq 1$ is satisfied (this rules out the $p=0$ case),
we will show that the
scheme is still unstable. Consider as initial data
a small perturbation of the form
\begin{equation}
{u}^0_0 = (0, 2\pi\epsilon )^T \;\;\; , 
\;\;\; u^0_k = (0,0)^T \mbox{ otherwise} \label{error}
\end{equation}
with $\epsilon $ arbitrary small. In Fourier space we have $\hat{u}^0(\xi) =
\epsilon(0,\, 1)^T$ for all $\xi$. The solution at time
$t=n\Delta t$ is, then, $\hat{u}^n = 
\epsilon (a^{n-1}nb,\, a^n)^T$. We will show that its norm grows
without bound when the number of gridpoints is increased (while
keeping the time and Courant factor fixed). We first notice that
($n_2\equiv 2n-2$),
\begin{equation}
\frac{\| u^n \|^2}{\| u^0 \|^2} = \int_{-\pi}^{\pi} \!\!\!\!
\left( n^2 | a|^{n_2} |b|^2 + |a|^{2n} \right)\, {d\xi  \over 2\pi }
\geq n^2 \int_{-\pi}^{\pi}  | a|^{
n_2} |b|^2  {d\xi \over 2\pi }. \label{bound}
\end{equation}
Expressions $|a|^2$ and $|b|^2$ are analytic functions of $\xi$ with
Taylor expansion $ |b|^2 = \lambda ^2 [\xi^2 + O(\xi^4)]\, $ and $
|a|^2 = 1 - [2\tilde{\sigma}\lambda+d(c\lambda)^4] \xi^4 + O(\xi^6) $,
with $d= -1$ for $p=1$, $d=1$ for $p=2$, and $d=0$ otherwise. From
these expansions one can see that, for all $p$, there are positive
constants $\alpha $ and $\rho$ such that
\begin{equation}
|a(\xi)|^2 \ge 1-\alpha \xi^4 \ge 0 \;\;\; , \;\;\;
|b(\xi)|^2 \ge \rho^2 \xi^2 \ge 0 \label{bound_b} \; , 
\end{equation}
for small enough $\xi$, say $|\xi |\leq \epsilon_1$.  
For $n>1$ and small enough $\xi$, say
$|\xi |\leq \epsilon _2$,
\begin{equation}
|a(\xi)|^{2n} \ge 1 -n\alpha \xi^4 \ge 0 \label{bound_a} \; ,
\end{equation}
where the last inequality holds for $|\xi | \leq (n\alpha)^{-1/4}$,
provided that $n$ is large enough so that $(n\alpha)^{-1/4}\leq
\min(\epsilon _1, \epsilon _2)$. Using bounds
(\ref{bound_b},\ref{bound_a}) and the 
VNC in inequality (\ref{bound}) gives
$$
\frac{\| u^n \|^2}{\| u^0 \|^2} \ge n^2
\int_{-(n\alpha)^{-1/4}}^{(n\alpha)^{-1/4}}\!\!\!\!(1-n\alpha\xi^4)\rho^2
\xi^2 \,\frac{d\xi}{2\pi} = \frac{4}{21\pi}
\frac{\rho^2}{\alpha^{3/4}} n^{5/4}
$$
which diverges for $n\rightarrow \infty$. This shows that in the
WH case the ICN and 2RK schemes 
are unstable, independent of the amount of dissipation. In fact, by examining
$|a(\xi=\pi)|$, one can see that  a necessary condition for the VNC (for any 
number of iterations) is 
$0 \le \tilde{\sigma }\lambda \le 1/8$.
Adding too much dissipation violates the VNC and the 
instability becomes worse.

Perturbations like that of (\ref{error}) are to be expected in a
numerical simulation due to truncation or roundoff errors. For high
enough resolution, such a perturbation will be amplified without
bound, spoiling any convergence. One expects that in the non-constant
coefficient or in the non-linear case the rate of growth of the
instability with number of gridpoints will be even faster (see example
in page 216 of \cite{kreiss}), as in figure \ref{Fourier}.


\begin{thebibliography}{99}

\bibitem{ADM} R. Arnowitt, S. Deser, C. Misner, in 
{\it Gravitation: An Introduction
to Current Research}, L. Witten, ed. (John Wiley, New York, 1962) 

\bibitem{BSSN} T.~W.~Baumgarte and S.~L.~Shapiro,
Phys.\ Rev.\ D {\bf 59}, 024007 (1999); M.~Shibata and T.~Nakamura,
Phys.\ Rev.\ D {\bf 52} (1995) 5428.

\bibitem{ICN} 
S.~A.~Teukolsky,
Phys.\ Rev.\ D {\bf 61}, 087501 (2000)

\bibitem{kreiss} B. Gustafsson, H.O. Kreiss, J. Oliger, {\em Time
dependent problems and difference methods} (Wiley, New York, 1995).

\bibitem{ec} 
A.~Anderson and J.~W.~York,
Phys.\ Rev.\ Lett.\  {\bf 81}, 1154 (1998)

\bibitem{stability} G. Calabrese, J. Pullin, O. Sarbach, and
M. Tiglio, gr-qc/0205073.

\bibitem{kst} L.~E.~Kidder, M.~A.~Scheel and S.~A.~Teukolsky,
Phys.\ Rev.\ D {\bf 64}, 064017 (2001).

\bibitem{tadmor} E. Tadmor, {\it The unconditional instability of
inflow-dependent boundary conditions in difference approximations to
hyperbolic systems}, Math. of Computation, {\bf 41}, 309 (1983).

\end{thebibliography}
\end{document}